\documentclass[twocolumn,trackchanges]{aastex62}
\usepackage[section]{placeins}
\usepackage{longtable}
\usepackage{hyperref}
\usepackage{url}
\usepackage{amsfonts}
\usepackage{amsmath}
\usepackage{amssymb}
\usepackage{pifont}
\usepackage{comment}
\graphicspath{{./}{figs/}}

\usepackage{enumitem}
\usepackage{fontawesome}

\newcommand{\xmark}{\ding{53}}

\defcitealias{MacLeod16}{M16}
\newcommand{\MM}{\citetalias{MacLeod16}}

\usepackage[hang,flushmargin]{footmisc} 
\usepackage{url}

\newcommand{\STScI}{\affiliation{Space Telescope Science Institute, 3700 San Martin Dr, Baltimore, MD 21218, USA}}

\newcommand{\JHU}{\affiliation{Physics and Astronomy Department, Johns Hopkins University, Baltimore, MD 21218, USA}}

\shorttitle{FLEET-TDE}
\shortauthors{Gomez et al.}

\begin{document}

\title{The Search for Thermonuclear Transients from the Tidal Disruption of a White Dwarf by an Intermediate Mass Black Hole}

\correspondingauthor{Sebastian Gomez}
\email{sgomez@stsci.edu}

\author[0000-0001-6395-6702]{Sebastian Gomez}
\STScI

\author[0000-0003-3703-5154]{Suvi Gezari}
\STScI\JHU

\begin{abstract}

The close encounter of a white dwarf (WD) with a black hole (BH) could result in the tidal disruption of the WD. During this encounter, the WD can undergo a thermonuclear explosion due to its tidal compression, resulting in an optical transient similar to a Type Ia supernova (SN Ia), hereafter a Ia-TDE. Nevertheless, this will only be physically observable if the BH is $\lesssim 10^5$M$_\odot$. Finding a Ia-TDE would therefore imply the discovery of an intermediate mass black hole (IMBH) $\lesssim 10^5$ M$_\odot$. Here, we search the entire Zwicky Transient Facility (ZTF) alert stream for the elusive Ia-TDEs. We restrict our search to nuclear transients in dwarf galaxies, the likely sites for IMBHs, and find a total of six possible nuclear Ia-TDE candidates. We find SN\,2020lrt to be the most likely Ia-TDE candidate, thanks to its strong resemblance to light curve and spectroscopic models of Ia-TDEs. We measure the stellar masses of the dwarf galaxies hosting these transients to be $\lesssim 10^{9}$M$_\odot$; if confirmed to harbor BHs, these would prove the existence of IMBHs in some of the lowest-mass galaxies known. Additionally, we searched for off-nuclear Ia-TDEs, but were unable to find more robust candidates in the outskirts of galaxies than in their nuclei. This supports the hypothesis that the nuclear Ia-TDEs candidates are WDs tidally compressed by IMBHs in the cores of galaxies, as opposed to a class of transient that can happen anywhere in a galaxy. We have laid the groundwork to systematically search for Ia-TDE candidates in existing and future time-domain surveys. Rapid characterization of their nature will result in not only the confirmation of a Ia-TDE, but also the unambiguous discovery of bonafide IMBHs.

\end{abstract}


\section{Introduction}\label{sec:intro}

Tidal disruption events (TDEs) occur when a star gets torn apart by the tidal forces of a supermassive black hole \citep{hills75,Rees88}. Following the disruption, about half of the stellar debris is expected to return towards the black hole and circularize into an accretion disk, beginning a phase of accretion when a bright optical transient can be observed \citep{Gezari09, Guillochon09}. At present, about 70 TDEs have been discovered across the electromagnetic spectrum (e.g., \citealt{Auchettl17, Mockler19, Velzen20, Velzen20_TDEs, Gezari21, Hammerstein22, Nicholl22}).

The TDEs known to date involve the disruption of non-degenerate stars, but TDEs of white dwarfs (WDs) are also expected to exist (e.g., \citealt{Luminet89, Rosswog09, Clausen11, Haas12, Cheng14, MacLeod16, Tanikawa17, Kawana18, Tanikawa18, Vick17, Maguire20, Chen22, Lam22}). When a WD is disrupted in a deep encounter with a black hole (BH), i.e., well within the tidal disruption radius, the WD can be tidally compressed enough to trigger thermonuclear ignition \citep{Luminet85}. The observable transient that results from this thermonuclear explosion is expected to appear similar, but less luminous than a Type Ia supernova (SN Ia), given that the disruption of a WD does not require it to have a mass close to the Chandrasekhar limit \citep{Rosswog08}. Throughout this paper, we will refer to this tidally-induced thermonuclear transient as a Ia-TDE.

In order for a WD to get tidally disrupted, the disrupting BH needs to be $\lesssim 10^5$ M$_\odot$ to prevent the WD from simply entering the BH \citep{Kobayashi04, Rosswog09, Gezari21}. Equation~\ref{eq:mbh} (taken from \citealt{Rosswog09}) defines the upper limit on the mass of the BH that disrupts the WD outside the event horizon of the BH:
\begin{equation}
M_{\rm BH} \lesssim 2.5 \times 10^5 \left(\frac{R_{\rm WD}}{10^9 {\rm cm}} \right)^{3/2}   \left(\frac{M_{\rm WD}}{0.6 M_\odot} \right)^{-1/2},
\label{eq:mbh}
\end{equation}
where $R_{\rm WD}$ and $M_{\rm WD}$ are the radius and mass of the WD, respectively. This mass range corresponds to the population of intermediate mass black holes (IMBHs), which span a range of $\approx 100 - 10^5$ M$_\odot$ \citep{Greene20}. These are rare discoveries, and to date, only a handful of non-stellar black holes below $\lesssim 10^5$ M$_\odot$ have been suggested to exist (e.g. \citealt{Kaaret01,Filippenko03,Farrell09,Bosch10,Reines13,Baldassare15,Pasham15,Mezcua15,Nguyen19,Baldassare20,Fusco22,Ward22}). Therefore, finding a Ia-TDE would be of great relevance to the study of BHs, since it would also indicate the unambiguous discovery of an IMBH. Furthermore, Ia-TDEs are expected to be multi-messenger sources detectable in gravitational waves by future space-based observatories such as the Laser Interferometer Space Antenna (LISA; \citealt{Amaro-Seoane17}) from nearby IMBHs residing in globular clusters or dwarf galaxies within the Local Group \citep{Rosswog09}.

The rates of WD TDEs are determined from dynamical models to be $\sim 10^{-6}$ per year per BH in dwarf galaxies \citep{MacLeod16}, with an order of magnitude lower rate expected from BHs in globular clusters \citep{Ramirez09}. While only a fraction of WD TDEs will penetrate deep enough into the tidal radius of a BH to trigger a thermonuclear explosion, a recent dynamical calculation by the Monte Carlo Cluster Simulator (MOCCA) survey suggests that the rate of WD TDEs from IMBHs in star clusters in the local Universe is high enough that the subsample of WD TDEs with a thermonuclear transient (Ia-TDEs) could be detected at the rate of $\sim 100-500$ per year by {\it Rubin} \citep{Tanikawa22}. So far, only a handful of Ia-TDE candidates have been reported in the literature. Swift J1644+57 was suggested to be one due to multiple recurring hard X-ray flares, potentially attributed to repeated partial disruptions of a gravitationally bound WD \citep{Krolik11}. XRT 000519 was a bright X-ray flash in M86, proposed to be either the disruption of a WD or a foreground accreting neutron star or asteroid \citep{Jonker13}. The gamma-ray burst (GRB) GRB060218 associated with the supernova 2006aj has X-ray spectra consistent with the disruption of a WD \citep{Shcherbakov13}, but was previously suggested to be powered by a relativistic jet \citep{Soderberg06}. The X-ray transients XT1 and XT2 were suggested to be tidal disruptions of WDs due to their short X-ray flare durations followed by a decay slope consistent with a WD disruption \citep{Peng19}. The bright optical transient AT\,2018cow was suggested to be a Ia-TDE by \cite{Kuin19} due to its X-ray variability and blackbody-like UV spectrum. Nevertheless, the nature of AT\,2018cow is highly uncertain and has been suggested to be anything from a relativistic jet within a fallback supernova \citep{Perley19}, an engine-powered transient \citep{Ho20}, or an interaction-powered fast blue optical transient \citep{Fox19}. Finally, calcium-rich gap transients have also been suggested to be the result of tidally compressed WDs, despite the fact that these occur in the outskirts of galaxies \citep{Sell15}.

In this study, we perform a systematic search for Ia-TDEs by searching the transient alert stream of the Zwicky Transient Facility (ZTF; \citealt{Bellm19}) and comparing the light curves and spectra of these transients to the theoretical model predictions of Ia-TDEs from \cite{MacLeod16}, hereafter \MM. Our search results in a sample of six nuclear candidates: Two were previously classified as SNe Iax, one as a SN II, and one as a SN Ib; the remaining two candidates do not have a spectroscopic classification. Despite the fact that none of these are confirmed to be Ia-TDEs, they all have photometric and/or spectroscopic features that are largely consistent with the Ia-TDE models from \MM. The most likely candidate, SN\,2021lrt, provides the best match to the optical light curve and spectroscopic models from \MM.

As an additional test for completeness, we present a sample of four, less likely, off-nuclear Ia-TDE candidates recovered using the same procedures used to find the six more likely nuclear Ia-TDEs. This reiterates that the presence of the most likely Ia-TDE candidates appears to be correlated with the nuclei of galaxies, supporting the hypothesis that Ia-TDEs are caused by IMBHs in the cores of galaxies, as opposed to being a kind of transient that can happen anywhere in a galaxy.

The feature that makes Ia-TDE truly unique from other types of SNe Ia is the expectation of multi-wavelength emission \citep{MacLeod16}. This includes both X-ray emission from an accretion disk and afterglow radio emission from a jet \citep{MacLeod14}. These features should be identifiable within about a month to a year after disruption, making a follow-up survey with X-ray and radio telescopes a feasible approach to verify their nature.

We have set the groundwork to systematically search transient alert streams to find Ia-TDEs based on their color, rise time, and location within their host galaxy. This will prove to be of critical importance once larger surveys such as the Legacy Survey of Space and Time (LSST) on the Vera C.~Rubin Observatory (\textit{Rubin}; \citealt{Ivezic19}) or the \textit{Nancy Grace Roman Space Telescope} (\textit{Roman}; \citealt{Spergel15}) High Latitude Time Domain Survey (HLTDS; \citealt{Rose21}) begin in 2024 and 2027, respectively. These surveys will increase the current transient discovery rate by $\sim 2$ orders of magnitude, facilitating the discovery of rare classes of transients such as Ia-TDEs.

The structure of the paper is as follows: in \S\ref{sec:data} we outline the data sources used for this project, including a description of the theoretical comparison models used. In \S\ref{sec:selection} we describe our target selection procedures. In \S\ref{sec:candidates} we describe each nuclear Ia-TDE candidate in detail and in \S\ref{sec:off} each off-nuclear Ia-TDE candidate. In \S\ref{sec:spectra} we describe the spectral features of the nuclear candidates that have spectra and in \S\ref{sec:host} we present models of the host galaxies of these candidates. Finally, we present our discussion and conclusions in \S\ref{sec:discussion} and \S\ref{sec:conclusions}, respectively. Throughout this work we assume a flat $\Lambda$CDM cosmology with \mbox{$H_{0} = 69.3$ km s$^{-1}$ Mpc$^{-1}$}, $\Omega_{m} = 0.286$, and $\Omega_{\Lambda} = 0.712$ \citep{hinshaw13}.

\begin{figure}[h]
    \begin{center}
    \centering
    {\includegraphics[width=\columnwidth]{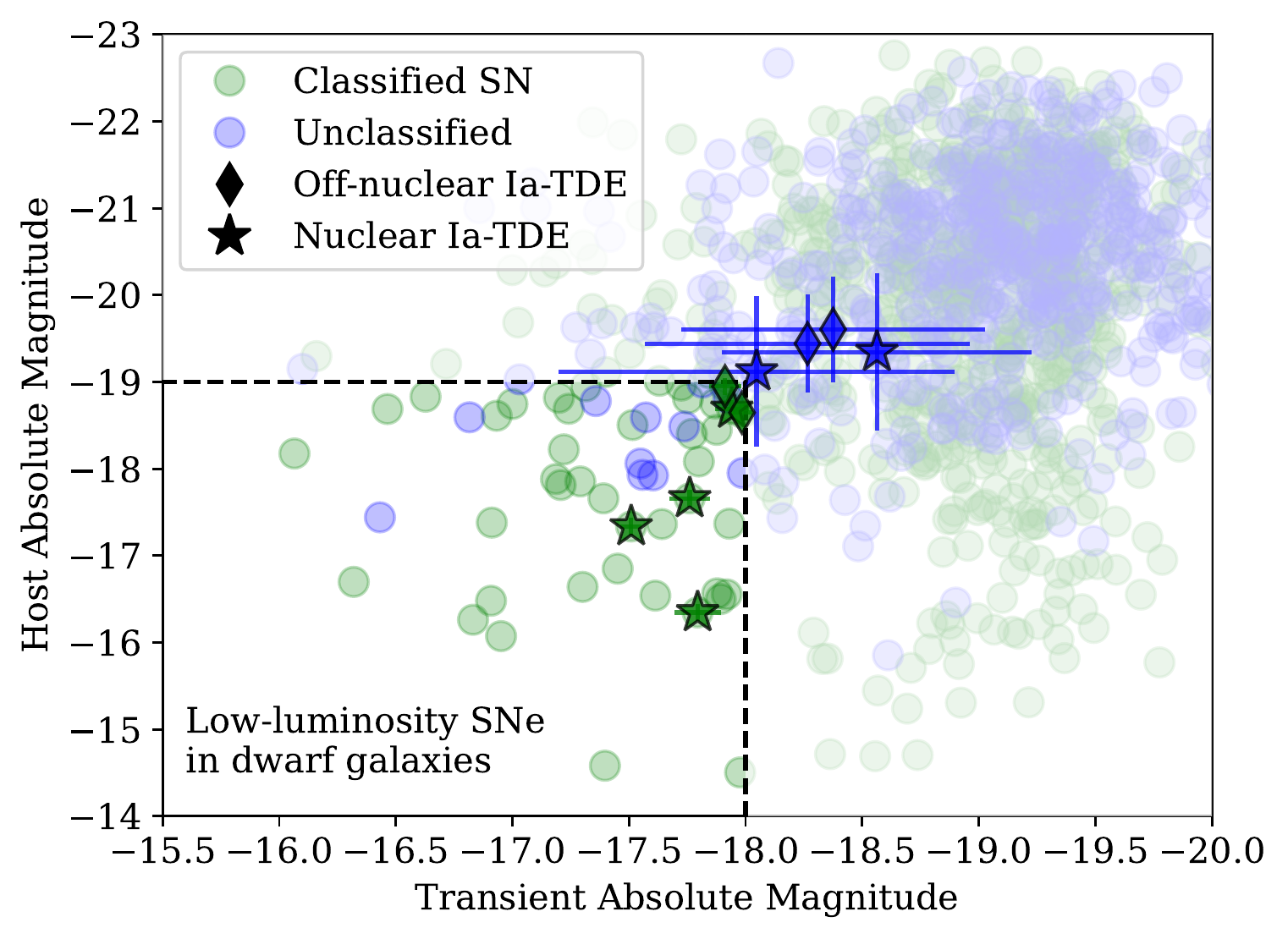}}
    \caption{$g$-band absolute magnitude of transients compared to the $g$-band absolute magnitude of their host galaxies. All data points are nuclear transients (813 classified in green and 630 unclassified in blue). The box on the lower left corner represents the intrinsically low-luminosity transients in low-luminosity host galaxies, where Ia-TDEs are expected to occur. There are 136 transients in the selection box, 46 classified and 90 unclassified. The star markers show the most likely nuclear Ia-TDE candidates, and the diamond markers show the less likely off-nuclear Ia-TDE candidates.
    \label{fig:absmags}}
    \end{center}
\end{figure}

\section{Data}\label{sec:data}

We focus our search for Ia-TDE candidates exclusively on transients detected by ZTF, given that this is the untargeted survey with the largest number of detected transients. We begin by downloading every transient from the Transient Name Server (TNS)\footnote{\label{ref:tns}\url{https://www.wis-tns.org/}}. As of 2022 Nov 9, there were a total of 82,976 transients on the TNS that happened on or after 2018, are above a declination of $-32$ deg, and are not classified as FRBs. These cuts are implemented to restrict the list to only transients that could be in the ZTF survey.

We then use the infrastructure developed for the FLEET algorithm \citep{Gomez20} to gather the light curves and host galaxy information for all these transients. We obtain ZTF light curves from the Automatic Learning for the Rapid Classification of Events (ALeRCE) broker \citep{Forster20}. We query a 1$\arcmin$ region around every transient from the PS1/$3\pi$ \citep{Chambers18} and Sloan Digital Sky Survey (SDSS) catalogs \citep{Alam15,Ahumada19} to obtain information about the host galaxy of each transient. Finally, we correct all photometry for Galactic extinction using the \cite{Schlafly11} dust maps to estimate $E(B-V)$ and the \cite{Barbary16} implementation of the \cite{Cardelli89} extinction law, assuming $R_{\rm V} = 3.1$.

For the most likely Ia-TDE candidates (described in \S\ref{sec:candidates}), we download additional ancillary data to help constrain their parameters. These include spectral observations from the TNS for the transients that have this data available, light curves from the ZTF forced-photometry service \citep{Masci19}, and additional host photometry and locations from the Gaia Data Release 3 catalog \citep{Gaia16, Gaia21}, the Wide-field Infrared Survey Explorer (WISE; \citealt{WISE}) catalog, and the Two Micron All Sky Survey (2MASS, \citealt{2MASS}) catalog.

For comparison purposes, we include a sample of well-observed SNe Ia. We select these from spectroscopically classified SNe in the TNS and download their ZTF light curves using ALeRCE. We restrict the sample to only the 1,000 best observed SNe Ia, which only includes SNe with at least 10 $g$- and 10 $r$-band detections, with at least 4 detections before peak and at least 6 after peak, and the latest detection being at least 10 days after peak in both $g$- and $r$-band. This sample includes a set of seven SNe Iax, which are a known subclass of SNe Ia which mimic some of the photometric and spectroscopic properties of Ia-TDEs (see \S\ref{sec:SNIa}). Given the limited sample size of SNe Iax, we relax the criteria and include any SNe Iax with at least 1 detection before peak instead of 4.

\begin{figure}
    \begin{center}
    \centering
    {\includegraphics[width=\columnwidth]{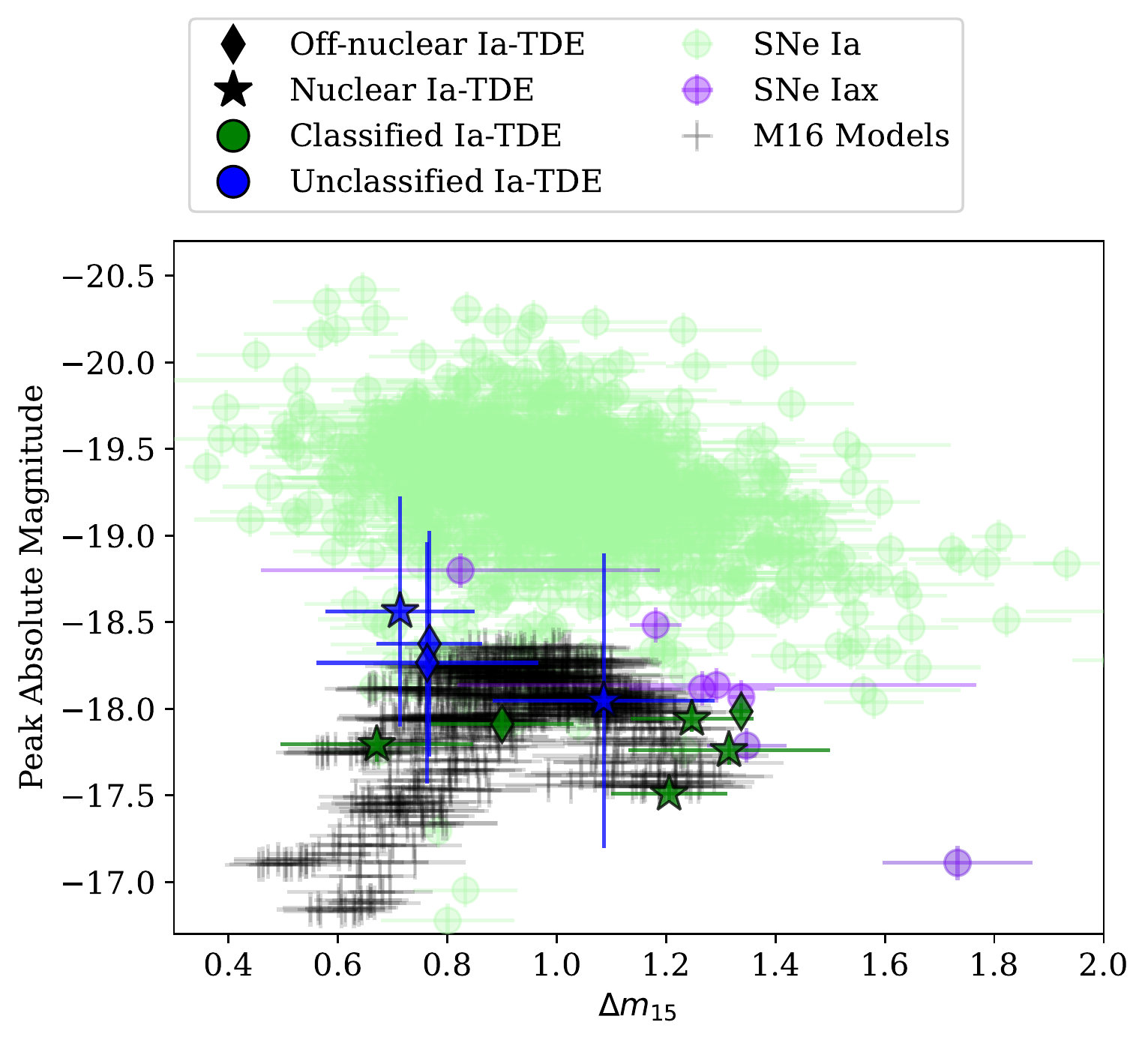}}
    \caption{The peak $r$-band absolute magnitude as a function of $\Delta m_{15}$ for the Ia-TDE candidates, the \MM\ models, and a sample of SNe Ia and SNe Iax. The stars denote the best nuclear Ia-TDE candidates, while the diamonds are for the off-nuclear Ia-TDE candidates. \label{fig:Ia_compare}}
    \end{center}
\end{figure}

\subsection{Comparison Model}\label{sec:model}

We compare the photometry and spectroscopy of Ia-TDE candidates to the theoretical light curve and spectral models from \MM. These models come from hydrodynamic simulations of the thermonuclear transient that results from the tidal compression of a single $0.6$ M$_\odot$ CO-WD disrupted by a $500$ M$_\odot$ BH. The WD is assumed to have a temperature of $5 \times 10^4$ K and be composed of 50\% Carbon and 50\% Oxygen. The simulation shows $0.13$ M$_\odot$ of iron-group elements synthesized in the explosion, which drives the strong similarity of these models to Type I SNe.

The light curves and spectra derived from these simulations were generated by \MM\ using the time-dependent radiation transport code {\tt SEDONA} \citep{Kasen06}. Orientation effects are also taken into account, provided in 30 angular bins equally spaced across $\cos(\theta)$ and $\phi$, the standard polar and azimuthal angles of spherical geometry.

\section{Candidate Selection}\label{sec:selection}

\begin{figure}
    \begin{center}
    \centering
    {\includegraphics[width=\columnwidth]{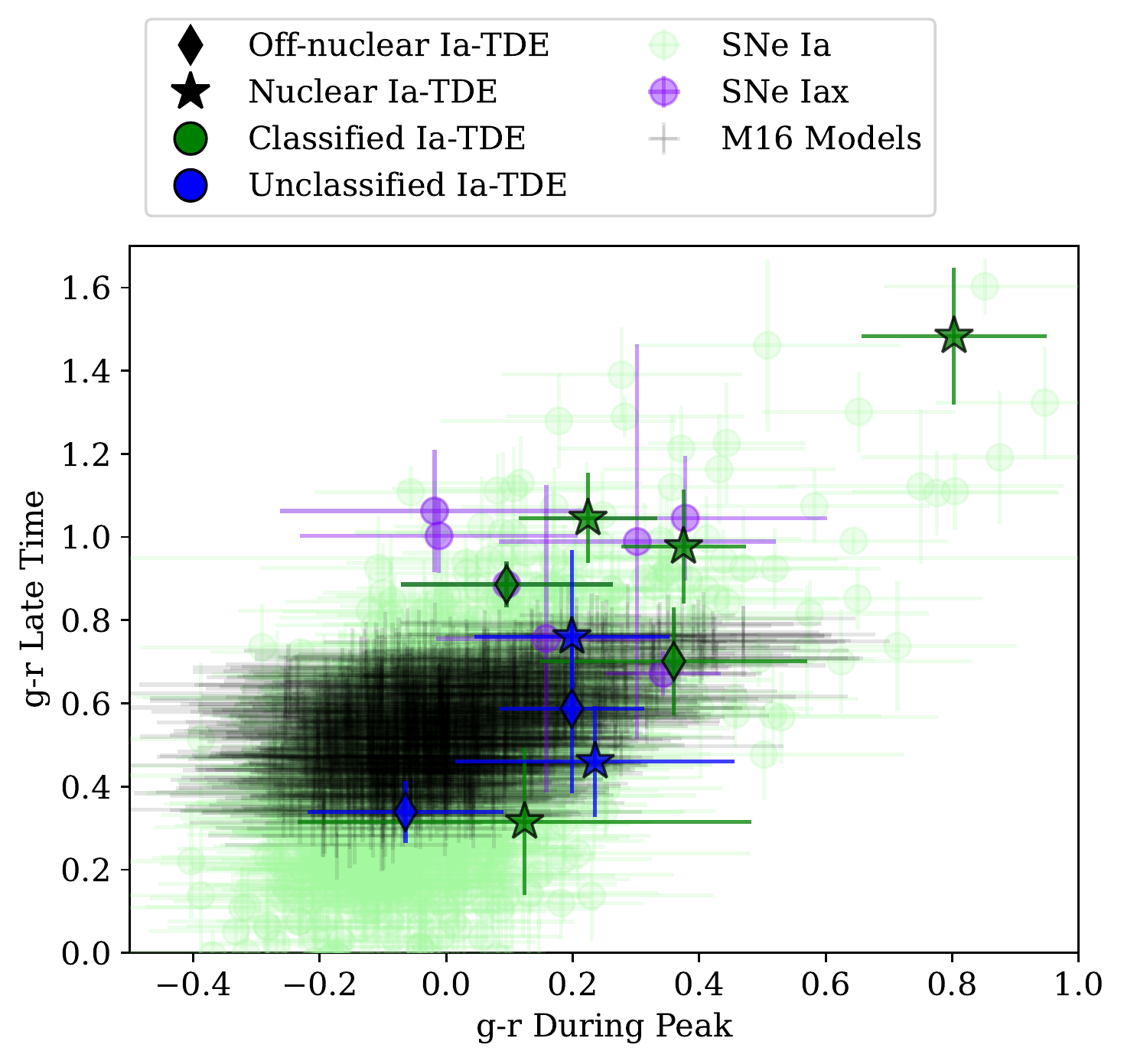}}
    \caption{$g-r$ color measured during peak and at 15 days post peak for the Ia-TDE candidates, the models from \MM\, and a sample of SNe Ia and SNe Iax. The stars denote the best nuclear Ia-TDE candidates, while the diamonds are for the off-nuclear Ia-TDE candidates. \label{fig:color_rise}}
    \end{center}
\end{figure}

Of the initial list of 82,976 transients considered for this work, we restrict the sample to the 28,895 transients that have at least 10 detections in any band in ZTF. Of those, 7,228 have a spectroscopic classification, and 21,667 remain unclassified. Since we are searching for Ia-TDEs, we restrict the sample in four ways: the transient must be nuclear, have a well-defined light curve, have an intrinsically low luminosity, and reside in a dwarf galaxy. The selection criteria for each point are described below.

For any given transient, the typical scatter in its position centroid in ZTF is $\sim 0.25 \arcsec$. Therefore, we adopt a very conservative restriction that the transient must be within 1.5$\arcsec$ of the nucleus of its host galaxy. The host galaxy of a transient is determined to be the closest object in SDSS or PS1/$3\pi$ with a probability of being a galaxy $\geq 0.1$, an optimal threshold determined in \cite{Gomez20} to rule out stars as possible contaminants. We further restrict the normalized host separation to be $R_n \leq 1$, where $R_n$ is defined as the transient-host separation divided by the half-light radius of the host. Finally, we restrict the sample to only include transients where the host is brighter than $m_r = 22$ mag to be detected in PS1/$3\pi$. These cuts reduce the sample to 1,838 classified and 5,728 unclassified transients.

We further restrict the candidates based on light curve quality, to be able to compare their light curves to models and measure their physical parameters. We only consider transients that have at least 1 $g$-band and 1 $r$-band data point during the rise, and at least 2 $g$-band and 2 $r$-band points during the decline, extending to at least 10 days after peak in both bands. We also remove any transients with long-term variability such as AGN or stellar flares. This is our most stringent cut, which reduced the sample to 813 classified transients and 630 unclassified ones. The values for the peak absolute magnitude and host absolute magnitude in $g$-band for all of these transients are shown in Figure~\ref{fig:absmags}.

The light curve models from \MM\ show that the brightest Ia-TDEs are expected to have a peak absolute magnitude of $M_g \sim -18$. Therefore, we restrict our search to only include transients that peak at $M_g \geq -18$, within $1\sigma$ of their photometric and redshift uncertainties. For the classified transients, we use the redshift reported to the TNS to calculate their peak absolute magnitude. For unclassified transients that do not have a host with a known spectroscopic redshift, we use their photometric redshift (photo-z) from SDSS and account for the uncertainty in the photo-z estimate before applying our magnitude selection cut.

In order for a WD to be disrupted by a BH, this BH has to be $\lesssim 10^5$ M$_\odot$, which are expected to reside in dwarf galaxies \citep{Kobayashi04, Rosswog09, East14, MacLeod14}. Given that the vast majority of galaxies in our selection sample do not have a known BH mass, we resort to using the absolute magnitude of the host galaxy as a proxy for galaxy mass, and therefore BH mass \citep{Danieli18}. For comparison, we use the galaxy mass measurements from the passive stellar evolution models of \cite{Maraston09} included in SDSS, and the \cite{Shen08} sample of BH mass estimates for $60,000$ quasars. We adopt a conservative threshold of $M_r \geq -19$ mag for the host galaxy. Galaxies dimmer than this have typical stellar masses below $\sim 10^{10}$ M$_\odot$, and their BHs are expected to be less than $\sim 10^7$ M$_\odot$ \citep{Robinson21}. We note that this is a very conservative cut and not our main discriminant. By imposing this cut we are only targeting to rule out black holes $\gtrsim 10^7$ M$_\odot$, or two orders of magnitude higher than the expected $10^5$ M$_\odot$ threshold.

Imposing these cuts on the transient peak absolute magnitude and host galaxy absolute magnitude reduced the list of candidates to 46 classified transients and 90 unclassified ones, shown in the bottom left corner of Figure~\ref{fig:absmags}. We note that some unclassified transients seem to lie well outside the selection area, but their uncertainties in photo-z are such that their magnitudes extend into the area being considered.

We fit the light curves of the 46 classified and 90 unclassified transients using the light curve model techniques from \cite{Hosseinzadeh20} and \cite{Villar20}, which is intended to be a generic non-physical parameterization of any SN light curve. From these models, we measure the $g-r$ color of the transient during peak, and 15 days after peak. Additionally, we use the light curve fits to derive an estimate of $\Delta m_{15}$, which is defined as the change in magnitude from peak to 15 days after peak. Finally, we calculate these same metrics for the light curve models of \MM, including all realizations of the models provided as a grid of observing angles $\theta$ and $\phi$ and place them at redshifts of $z = 0.001, 0.01$ and $0.1$ to provide representative observed parameters for these models.

We visually inspect the 46 classified and 90 unclassified low-luminosity transients in dwarf galaxies. The light curves of some candidates were either too slow or too fast to be consistent with the Ia-TDE models from \MM, despite having the right peak luminosity. Similarly, we rule out some transients that have a $g-r$ color that does not match the predictions from the \MM\ models. Finally, we inspect the spectra of all classified transients and reject any objects with spectra that were significantly different from the spectral models from \MM, mostly eliminating only Type II SNe. That leaves a final sample of four spectroscopically classified nuclear Ia-TDE candidates (SN\,2019uqp, SN\,2020lrt, SN\,2020bpf, SN\,2021jun) and two unclassified ones (AT\,2021hsd, AT\,2020djm). Each candidate is described in detail in \S\ref{sec:candidates}.

\begin{deluxetable*}{cccccccccc}
    \tablecaption{Ia-TDE Candidates Physical Parameters \label{tab:sne}}
    \tablewidth{0pt}
    \tablehead{\colhead{Name} & \colhead{Type} & Redshift & \colhead{SN mag} & \colhead{Host mag} & \colhead{Host Sep.} & \colhead{$R_n$} & \colhead{$\log(M_*)$} & Reference & $\chi^2$/d.o.f. \\
                              &                &          & $r$-band         & $r$-band           & arcsec                    &                 &  (M$_\odot$)   &      &}
        \startdata
	\multicolumn{9}{c}{Nuclear}   \\ 
        \hline
        2019uqp & SN II   &  0.040            & $-17.74 \pm 0.10$ & $-16.29$          & $0.32 \pm 0.35$ & $0.23 \pm 0.26$ & $ 7.72 \pm 0.11$ & [2]     & 5.5  \\
        2020bpf & SN Ib   &  0.027            & $-17.47 \pm 0.05$ & $-17.30$          & $0.47 \pm 0.25$ & $0.54 \pm 0.29$ & $ 6.95 \pm 0.07$ & [5]     & 39.1 \\
        2020lrt & SN Iax  &  0.046            & $-17.88 \pm 0.07$ & $-18.63$          & $0.58 \pm 0.59$ & $0.24 \pm 0.24$ & $ 8.59 \pm 0.05$ & [4]     & 9.9  \\
        2021jun & SN Iax  &  0.040            & $-17.70 \pm 0.09$ & $-17.60$          & $0.21 \pm 0.41$ & $0.12 \pm 0.24$ & $ 8.84 \pm 0.06$ & [6,7]   & 18.5 \\
        2021hsd & \nodata & $0.064 \pm 0.021$ & $-17.92 \pm 0.80$ & $-19.08 \pm 0.87$ & $0.55 \pm 0.36$ & $0.18 \pm 0.12$ & $ 8.96 \pm 0.03$ & \nodata & 5.9  \\
        2020djm & \nodata & $0.074 \pm 0.021$ & $-18.46 \pm 0.52$ & $-19.11 \pm 0.84$ & $0.07 \pm 0.28$ & $0.04 \pm 0.17$ & $ 8.26 \pm 0.02$ & \nodata & 3.8  \\
        \hline
	\multicolumn{9}{c}{Off-nuclear}   \\ 
        \hline
        2020sck & SN Iax  &  0.017            & $-17.98 \pm 0.05$ & $-18.65$          & $2.33 \pm 0.23$ & $0.71 \pm 0.07$ & \nodata          & [9]     & 174.3 \\
        2022lce & SN Ia   &  0.036            & $-17.91 \pm 0.07$ & $-18.95$          & $5.39 \pm 0.23$ & $1.65 \pm 0.07$ & \nodata          & [8]     & 43.7  \\
        2020xmo & \nodata & $0.071 \pm 0.019$ & $-18.38 \pm 0.65$ & $-19.61 \pm 0.61$ & $3.56 \pm 0.37$ & $1.16 \pm 0.12$ & \nodata          & \nodata & 16.4  \\
        2021xpf & \nodata & $0.069 \pm 0.018$ & $-18.26 \pm 0.70$ & $-19.44 \pm 0.56$ & $3.37 \pm 0.46$ & $1.74 \pm 0.24$ & \nodata          & \nodata & 6.7   \\
        \enddata
        \tablecomments{List of best Ia-TDE candidates. The transients with a spectroscopic class have a spectroscopic redshift from TNS, while the unclassified transients rely on photo-z measurements from SDSS with a corresponding uncertainty. We include the transient peak absolute magnitude in $r$-band, the host galaxy absolute magnitude in $r$-band, the separation between the transient and host galaxy, and normalized host separation $R_n$. The last column lists the $\chi^2$ per degree of freedom value for the best-fit \MM\ model to each light curve. 1: \cite{2022fwg}; 2: \cite{2019uqp}; 3: \cite{2018ldk}; 4: \cite{2020lrt}; 5: \cite{2020bpf}; 6: \cite{2021jun_a}; 7: \cite{2021jun_b}; 8: \cite{2022lce}; 9: \cite{2020sck}.}
\end{deluxetable*}

\begin{deluxetable*}{cccccccc}
    \tablecaption{Ia-TDE Candidates Features \label{tab:features}}
    \tablewidth{0pt}
    \tablehead{\colhead{Name} & \colhead{Peak Mag.} & \colhead{Peak $g-r$} & \colhead{Late $g-r$} & \colhead{Nuclear} & \colhead{Spectra} & \colhead{$\Delta m_{15}$} & \colhead{Galaxy Mass}}
        \startdata
	\multicolumn{8}{c}{Nuclear}   \\ 
        \hline
        2019uqp               & \checkmark          & \checkmark           & \checkmark           & \checkmark        & \xmark       & \checkmark  & \checkmark \\
        2020bpf               & \checkmark          & \checkmark           & $\sim$               & $\sim$            & \xmark       & \checkmark  & \checkmark \\
        2020lrt               & \checkmark          & \checkmark           & \checkmark           & \checkmark        & \checkmark   & \checkmark  & \checkmark \\
        2021jun               & \checkmark          & \xmark               & \xmark               & \checkmark        & \checkmark   & \checkmark  & \checkmark \\
        2021hsd               & \checkmark          & \checkmark           & \checkmark           & $\sim$            & \nodata      & \checkmark  & \checkmark \\
        2020djm               & \checkmark          & \checkmark           & \checkmark           & \checkmark        & \nodata      & \checkmark  & \checkmark \\
        \hline
	\multicolumn{8}{c}{Off-nuclear}   \\ 
        \hline
        2020sck               & \checkmark          & \checkmark           & \checkmark           & \xmark            & $\sim$       & \checkmark  & \nodata    \\
        2022lce               & \checkmark          & \checkmark           & \checkmark           & \xmark            & $\sim$       & \checkmark  & \nodata    \\
        2020xmo               & \checkmark          & \checkmark           & \checkmark           & \xmark            & \nodata      & \checkmark  & \nodata    \\
        2021xpf               & \checkmark          & \checkmark           & \checkmark           & \xmark            & \nodata      & \checkmark  & \nodata    \\
        \enddata
        \tablecomments{List of Ia-TDE candidates and whether or not their features are consistent with the Ia-TDE models from \MM. We assign a \checkmark\ if the feature is within less than $1\sigma$ of the model prediction, a\ $\sim$ if it is within $2\sigma$, and a \xmark\ if it is higher than that. A \nodata indicates no information is available. For the spectra, we assign a \checkmark\, $\sim$, or \xmark\ based on a qualitative comparison of the observed spectra to the models of \MM. We also include the best fit stellar mass for the host galaxy of the nuclear transients, for this parameter we assign a \checkmark\ if the mass is below $\log(M_*) < 9.7$.
        }
\end{deluxetable*}

\subsection{Comparison to SNe Ia} \label{sec:SNIa}

Ia-TDEs are expected to appear spectroscopically similar to SNe Ia, but with lower luminosity. This is reminiscent of SNe Iax, a class of SN characterized by low photospheric velocities of $\sim 2000 - 8000$ km s$^{-1}$ during peak, hot photospheres, and low peak luminosities between $-14.2 \gtrsim M_V \gtrsim -18.9$ \citep{Foley13, Jha17, Lee22}. One leading model to explain SNe Iax is a pure-deflagration explosion of a CO-WD, where the explosion is triggered by accretion of helium onto the WD, which does not necessarily fully disrupt the star \citep{Jha17}. It is likely that if Ia-TDEs are detected, they might appear classified as SNe Iax, as is the case for the nuclear Ia-TDE candidates SN\,2021jun, and SN\,2020lrt.

In Figure~\ref{fig:Ia_compare} we show the peak absolute magnitude and $\Delta m_{15}$ of the best Ia-TDE candidates, along with equivalent values for the \MM\ models, and a sample of SNe Ia and SNe Iax. The figure shows how all Ia-TDE candidates overlap with the region of parameter space spanned by the \MM\ models, but lie at a significantly lower luminosity than the population of normal SNe Ia. SNe Iax on the other hand appear to lie in between the Ia-TDE candidates and normal SNe Ia. In Figure~\ref{fig:color_rise} we show the $g-r$ color measured during peak and 15 days after peak for the same transients and models. All Ia-TDE candidates lie at most 2$\sigma$ away from the population of Ia-TDE models. One exception is SN\,2021jun, which appears to be much redder, but we nevertheless retain this candidate for comparison due to its strong spectral similarity to the \MM\ Ia-TDE models.

\vspace{-0.0cm}
\section{I\MakeLowercase{a}-TDE Candidates}\label{sec:candidates}

We compiled a total of six nuclear Ia-TDE candidates. These are transients located in the nuclei of dwarf galaxies, with intrinsically low luminosities, enough photometry to characterize their light curves, and good matches to the predictions from the \MM\ Ia-TDE models. Four of these candidates have spectroscopic classifications, and two do not. Since the \MM\ models are provided as a function of observing angle $\theta$ and $\phi$ (in 30 bins each), we find the orientation angles that best fit the light curve of each Ia-TDE candidate by doing a $\chi^2$ fit. We describe each candidate in detail below and provide a summary of all candidates in Tables~\ref{tab:sne} and \ref{tab:features}.

\begin{figure*}
    \begin{center}
    \centering
    {\includegraphics[width=2\columnwidth]{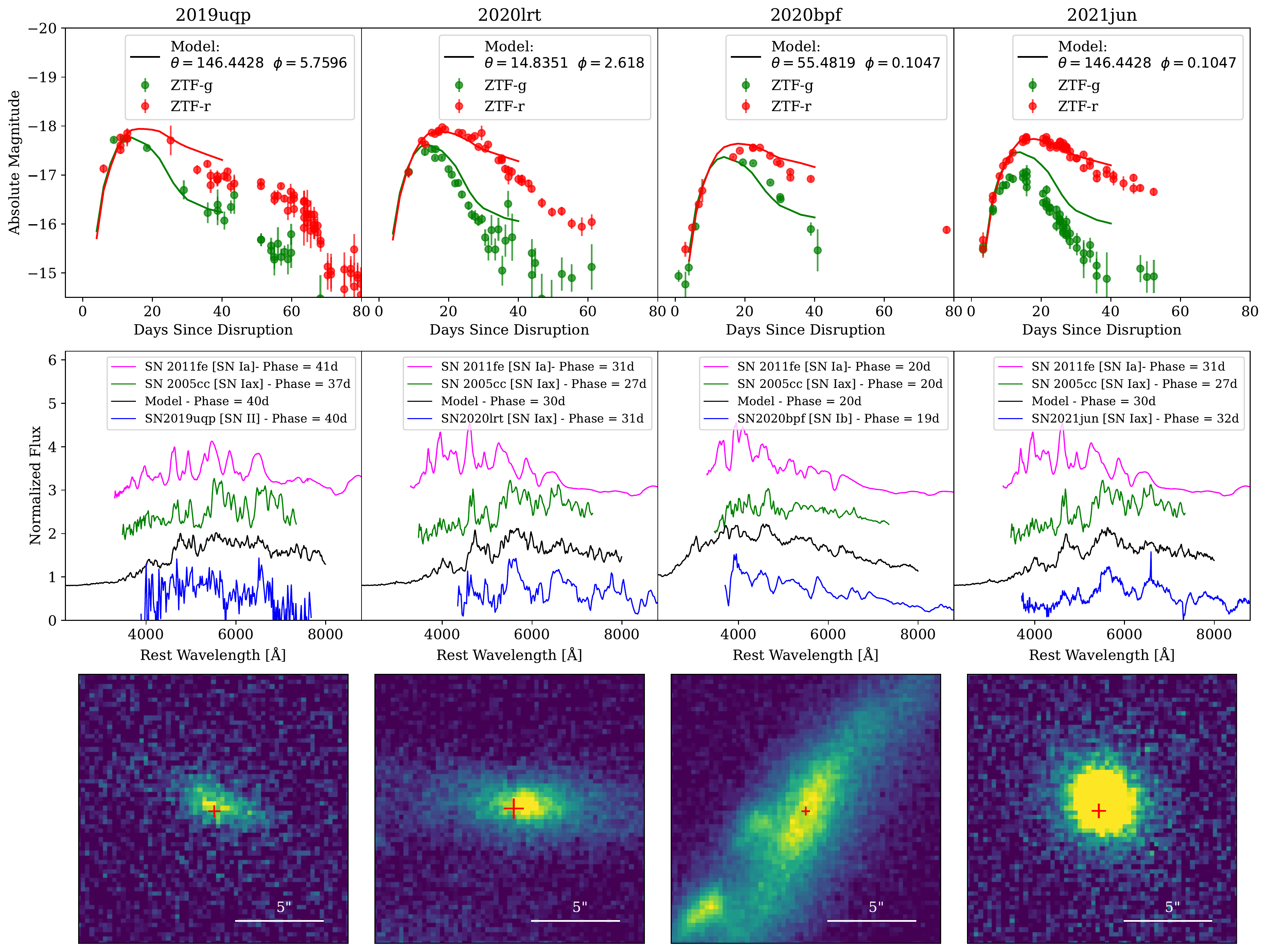}}
    \caption{\textit{Top row:} ZTF photometry in $g$- and $r$-band compared to the best fit Ia-TDE model from \MM, in terms of observing geometry angles $\theta$ and $\phi$. \textit{Middle row:} Spectra of the best Ia-TDE candidates, corresponding model from \MM, and a comparison SNe Ia and SN Iax from \cite{Nugent11} and \cite{Blondin12} at a similar phase to that of the candidate. \textit{Bottom row:} $r$-band PS1/$3\pi$ images of the host galaxies of the Ia-TDE candidates, the red cross marks the center of the transient and its size corresponds to the $3\sigma$ position uncertainties. \label{fig:classified}}
    \end{center}
\end{figure*}

\begin{figure}
    \begin{center}
    \centering
    {\includegraphics[width=1\columnwidth]{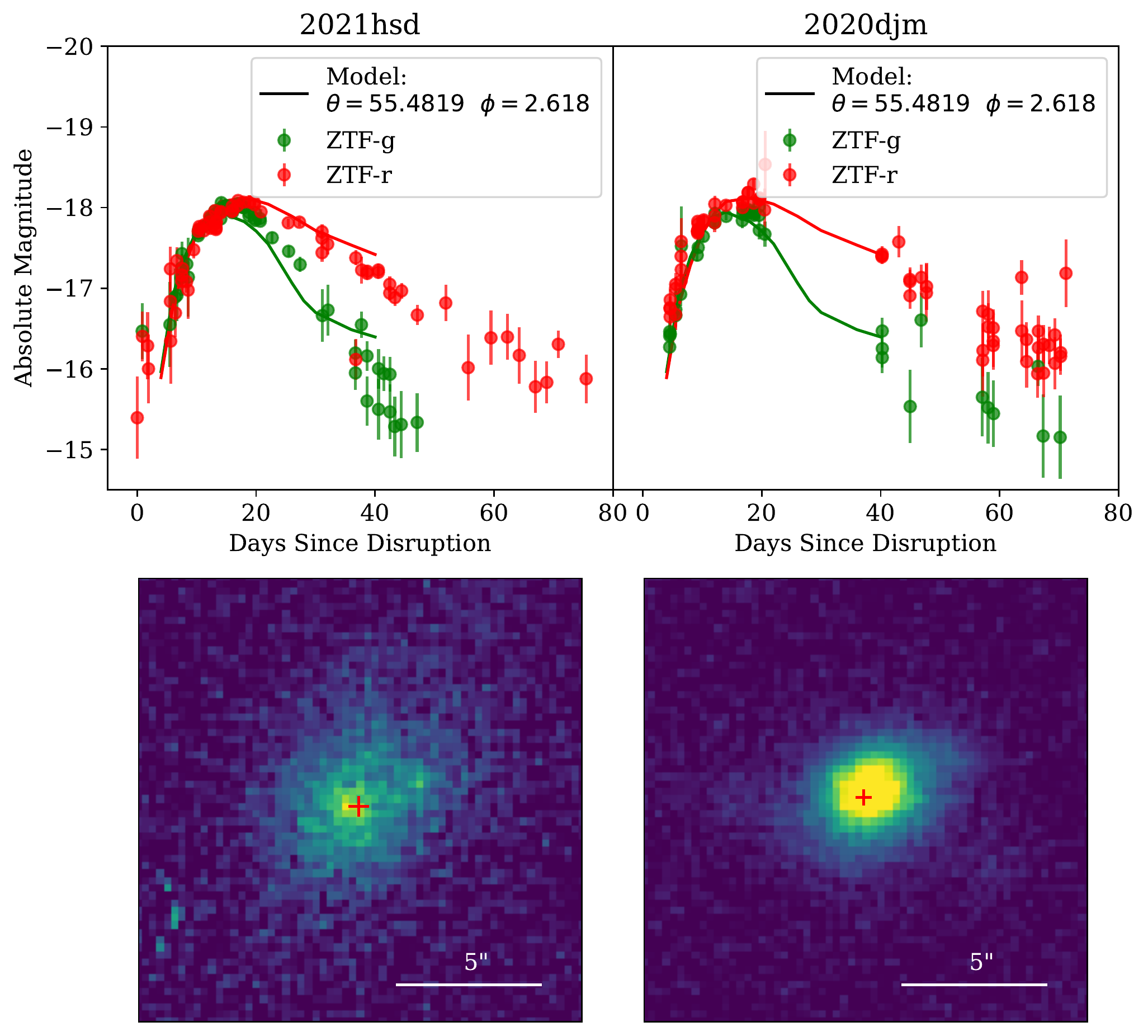}}
    \caption{\textit{Top row:} ZTF photometry in $g$- and $r$-band compared to the best fit Ia-TDE model from \MM, in terms of observing geometry angles $\theta$ and $\phi$. \textit{Bottom row:} $r$-band PS1/$3\pi$ images of the host galaxies of the Ia-TDE candidates, the red cross marks the center of the transient and its size corresponds to the $3\sigma$ position uncertainties. 
    \label{fig:unclassified}}
    \end{center}
\end{figure}

For the candidates with a spectroscopic classification, we adopt the redshift reported to the TNS as the redshift of the transient and host galaxy. Figure~\ref{fig:classified} shows all classified candidates, their light curves, spectra, and host galaxy image. For the two transients that do not have a spectroscopic classification or a spectroscopic redshift, we adopt the photo-z estimate from SDSS. We then find the redshift value that best fits the \MM\ models, as long as it is within $\pm 1 \sigma$ of the photo-z uncertainty. Figure~\ref{fig:unclassified} shows all unclassified candidates, their light curves, and host galaxy image.

\subsection{2019uqp}\label{sec:2019uqp}

SN\,2019uqp was classified as a SN II by \cite{2019uqp} at around 40 days after explosion. The light curve shape and color are consistent with a Ia-TDE. The spectra are too noisy to provide a definitive classification, but it appears consistent with a SN II. If this is a Ia-TDE, it is possible the hydrogen emission could be coming from interaction with pre-existing circumstellar material. The location of the transient is consistent with being nuclear within less than $1\sigma$.

\subsection{2020bpf}\label{sec:2020bpf}

SN\,2020bpf was classified as a SN Ib by \cite{2020bpf} at around 19 days after explosion. The light curve shape and color are consistent with a Ia-TDE, but the late-time color is slightly redder than what models predict. The spectra are a reasonable match to models of Ia-TDEs, but also match a SN Ib. The transient is consistent with being nuclear when compared to the host centroid of $3\pi$ within $2\sigma$. The elongated nature of the host galaxy makes the centroid determination uncertain.

\subsection{2020lrt}\label{sec:2020lrt}

SN\,2020lrt was classified as a SN Iax by \cite{2020lrt} at around 30 days after explosion. The light curve shape and color are consistent with a Ia-TDE. The spectra are a good match to what Ia-TDE models predict. We note the uncertainty on the centroid of the galaxy is large, but the transient is still consistent with being nuclear within $1\sigma$. This is the most likely Ia-TDE from our sample of candidates.

\subsection{2021jun}\label{sec:2021jun}

SN\,2021jun was classified as a SN Iax by \cite{2021jun_a} at around 30 days after explosion. The light curve shape is generally consistent with Ia-TDE models, but is much redder than expected. Nevertheless, the spectra are an excellent match to what Ia-TDE models predict. The centroid of the transient is within $2\sigma$ of the center of its host galaxy. 

\subsection{2021hsd}\label{sec:2021hsd}
The light curve of 2021hsd is consistent with the predictions from Ia-TDE models, but the $r$-band light curve fades faster than models predict. While the transient is only $1.5\sigma$ away from the center of its host galaxy, the location is still consistent with the very dense nucleus of the galaxy. We adopt a best-fit redshift of $z = 0.058$, or $0.3 \sigma$ lower than the quoted photo-z from SDSS.

\subsection{2020djm}\label{sec:2020djm}
The light curve of 2020djm is a great match to the predictions from Ia-TDE models from \MM. The transient is within $1\sigma$ of the host galaxy centroid from 3$\pi$. We adopt a best-fit redshift of $z = 0.0646$, or $0.45 \sigma$ lower than the quoted photo-z from SDSS.

\section{Off-Nuclear Transients}\label{sec:off}

Up to this point, we have focused our search for Ia-TDEs exclusively around the nuclei of galaxies, since we expect these to be the sites of IMBHs in the cores of galaxies. To test if the presence of good Ia-TDE candidates is correlated with the nuclei of galaxies, or whether it is simply a class of transient that can happen anywhere in a galaxy, we perform an identical search to that described in \S\ref{sec:selection}, but removing the nuclear constraint. In this case, we run through the exact same selection procedure as before and reach a total of 302 classified and 453 unclassified off-nuclear transients; compare this to the 46 classified and 90 unclassified nuclear transients. We exclude any transient that was selected as a candidate in \S\ref{sec:selection} to prevent duplication.

\begin{figure*}
    \begin{center}
    \centering
    {\includegraphics[width=2\columnwidth]{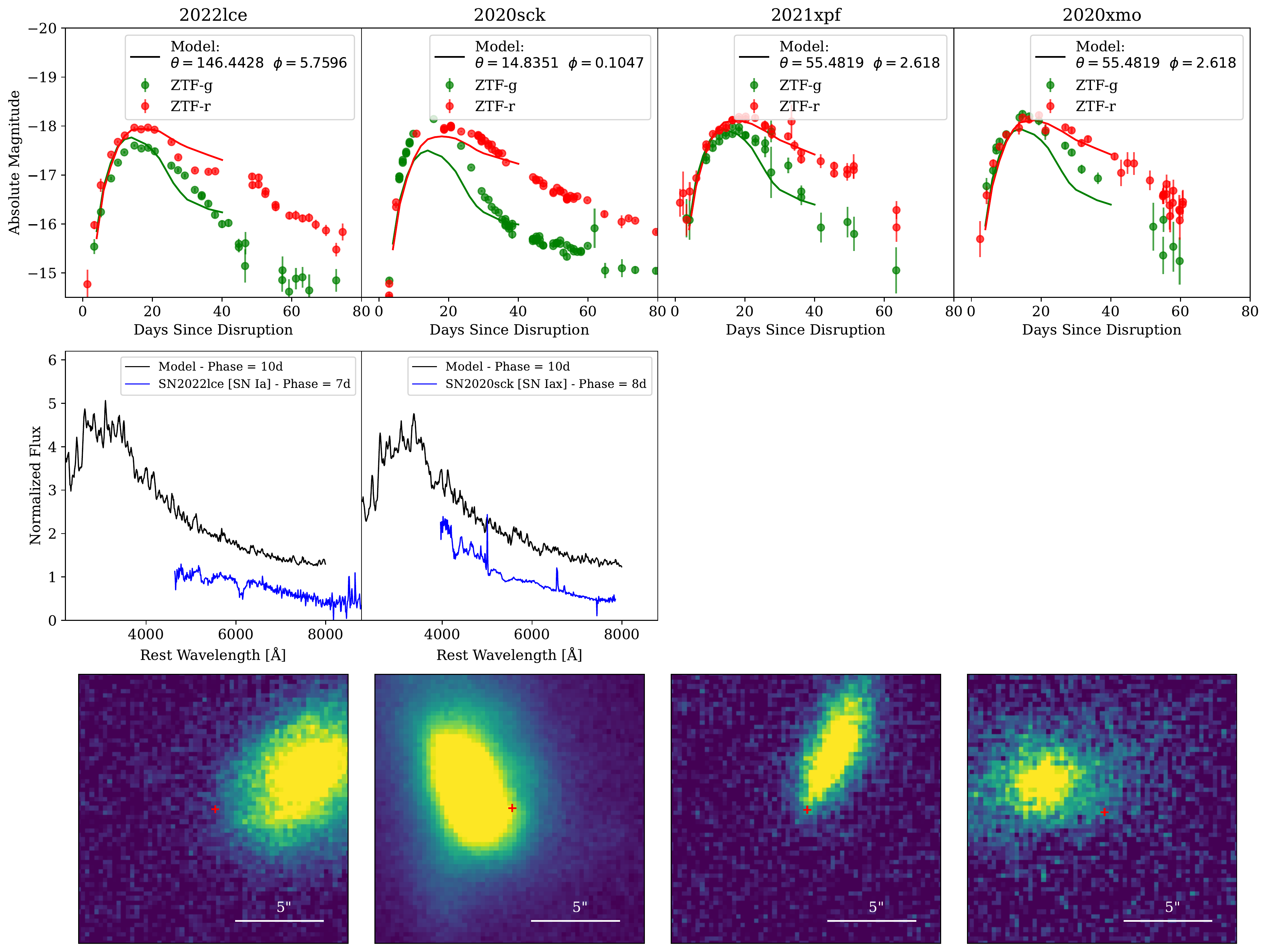}}
    \caption{Same as Figure~\ref{fig:classified}, but for the off-nuclear transients. 2021xpf and 2020xmo do not have a spectroscopic classification.
    \label{fig:off}}
    \end{center}
\end{figure*}

We find two classified and two unclassified off-nuclear transients that are reasonable matches to Ia-TDEs. This is comparable to the sample size of nuclear Ia-TDE candidates but derived from an initial list about an order of magnitude larger. This suggests that the presence of good Ia-TDE candidates is correlated with the nuclei of galaxies, as a much smaller fraction of Ia-TDE candidates appears to be off-nuclear. We describe each off-nuclear candidate in detail below. Figure~\ref{fig:off} shows all off-nuclear candidates, their light curves, spectra (when available), and host galaxy image. We provide a summary of all off-nuclear candidates in Table~\ref{tab:sne} and \ref{tab:features}.

\subsection{2020sck}\label{sec:2020sck}
SN\,2020sck was classified as a SN Iax by \cite{2020sck} at around 7 days after explosion. The light curve shape and color are consistent with a Ia-TDE. The spectrum is generally consistent with Ia-TDE models, but given that this was taken very early in the SN evolution, the continuum dominates the spectrum and a conclusive match is uncertain. The transient lies $2.3\arcsec$ away from its most likely host galaxy. We note that the peak absolute magnitude of SN\,2020sck is slightly higher than what the \MM\ models predict.

\subsection{2022lce}\label{sec:2022lce}
SN\,2022lce was classified as a SN Ia by \cite{2022lce} at around 8 days after explosion. The light curve shape and color are consistent with a Ia-TDE. The spectrum is generally consistent with Ia-TDE models, but given that this was taken very early in the SN evolution, the continuum dominates the spectrum and a conclusive match is uncertain. The transient lies $5.4\arcsec$ away from its most likely host galaxy. The late-time evolution of the transient appears to fade slower than what the \MM\ models predict.

\subsection{2020xmo}\label{sec:2020xmo}
The light curve of 2020xmo is consistent with the predictions from Ia-TDE models, but the light curve appears bluer than what models predict. The transient lies $3.6\arcsec$ away from its most likely host galaxy.

\subsection{2021xpf}\label{sec:2021xpf}
The light curve of 2020xmo is consistent with the predictions from Ia-TDE models. The transient lies $3.4\arcsec$ away from its most likely host galaxy.

\subsection{Summary}\label{sec:2021xpf}

\begin{figure}
    \begin{center}
    \centering
    {\includegraphics[width=\columnwidth]{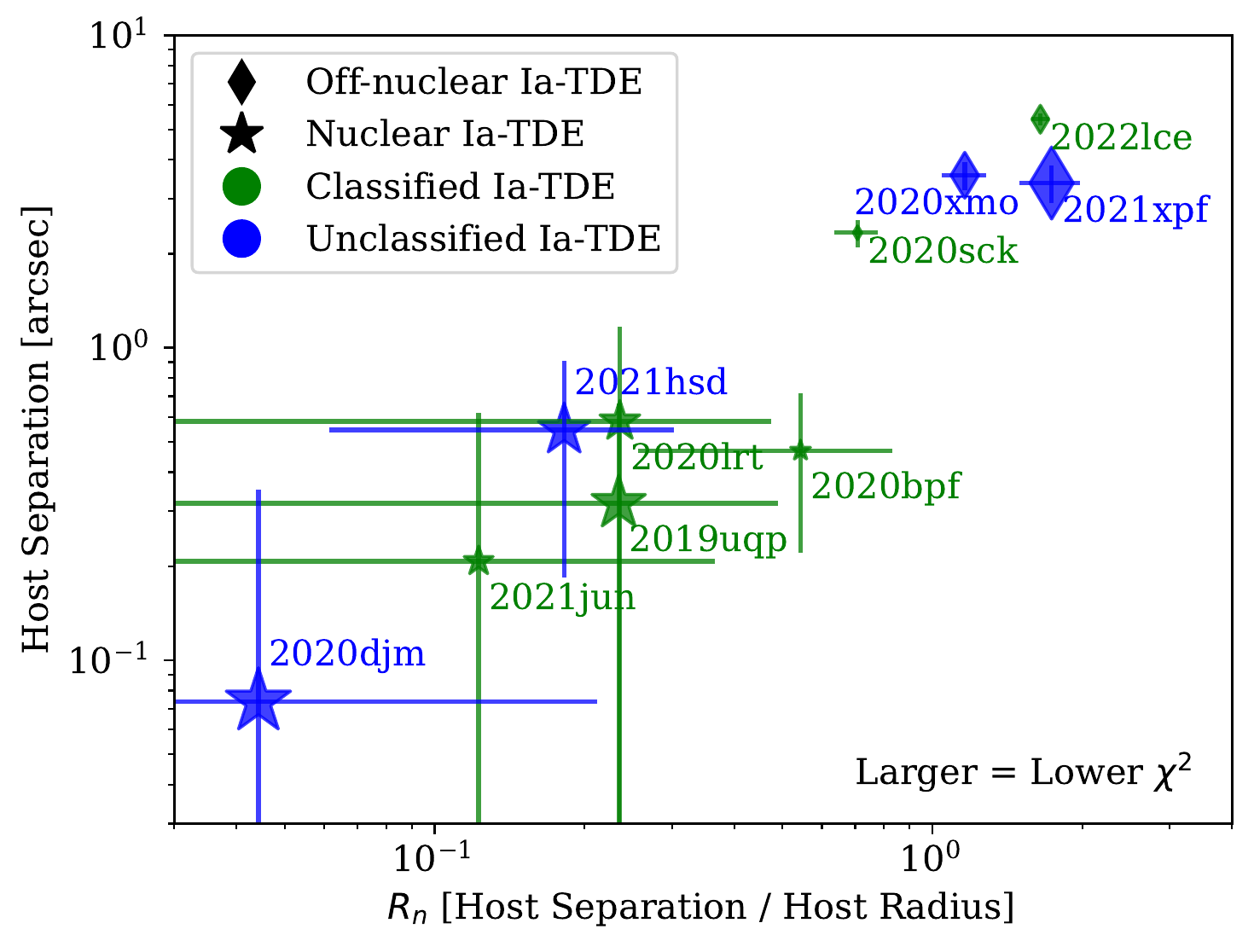}}
    \caption{Separation between transients and their host galaxy as a function of normalized host separation $R_n$. The nuclear Ia-TDE candidates, shown as stars, appear to be a better match to the Ia-TDE models from \MM\ than the best off-nuclear candidates, shown as diamonds. The relative size of each data point corresponds to $1 / \chi^2$ and represents the goodness of fit to the \MM\ light curve models.
    \label{fig:separations}}
    \end{center}
\end{figure}

In conclusion, we have found no significantly better Ia-TDE candidate in the off-nuclear sample of transients than in the corresponding nuclear sample. Supporting the theory that the six nuclear candidates are more likely to be Ia-TDEs. In Figure~\ref{fig:separations} we show the separation between the transients and their host galaxy as a function of normalized host separation $R_n$. There is a clear divide between the transients that are consistent with being nuclear and those that are clearly off-nuclear.

Alternatively, off-nuclear TDEs could be a complicating factor. Studies based on cosmological hydrodynamical simulations have suggested that off-center IMBHs could be common in dwarf galaxies \citep{Bellovary21}, with some observational evidence for offset AGN in dwarf galaxies from radio observations \citep{Reines20}. BHs could become off-center if they result from the merger of two black holes, depending on the velocity and parameters of the merger \citep{Blecha11}. Additionally, IMBHs could also exist in globular clusters (e.g., \citealt{Miller02, Haberle21, Vitral21}) in addition to the BH in the core of the galaxy.

\section{Spectroscopy}\label{sec:spectra}

\begin{figure}
    \begin{center}
    \centering
    {\includegraphics[width=\columnwidth]{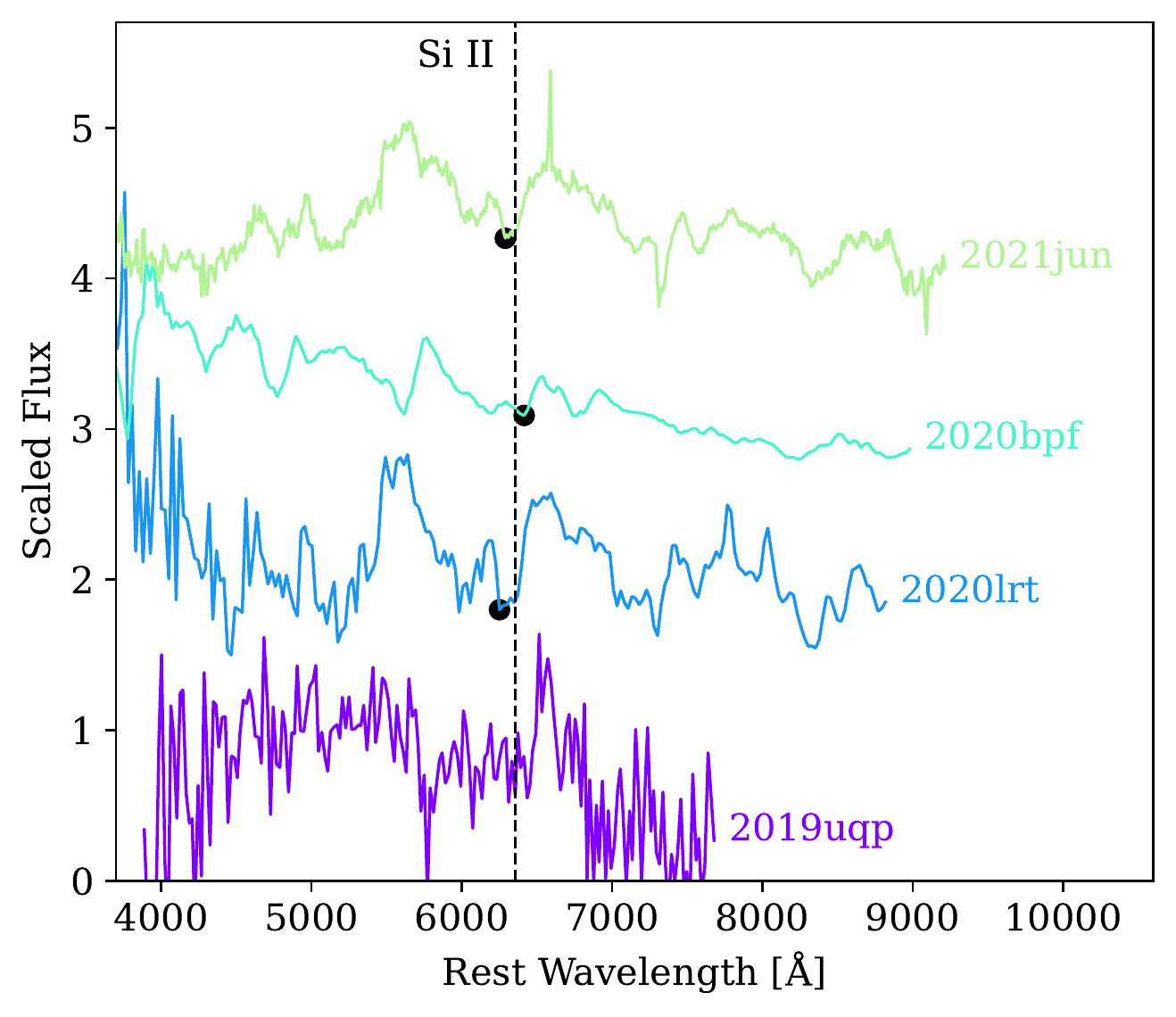}}
    \caption{Spectra of the four classified nuclear Ia-TDE candidates. The black dot shows the trough of the \ion{Si}{2} $\lambda$6355 line for each transient. Compare to Figure 9 of \MM. \label{fig:spectra}}
    \end{center}
\end{figure}

A key feature expected in Ia-TDEs is a large velocity offset of $\sim 10,000$ km s$^{-1}$ that results from Doppler shifts due to the high orbital motion of the ejecta imparted from the WD at the time of tidal compression and subsequent detonation. For the four classified transients, we attempt to fit the absorption component of \ion{Si}{2} with a Gaussian function to determine the velocity derived from the trough of this line, shown in Figure~\ref{fig:spectra}. We find the following velocities for each spectrum: $\sim - 4,900$ km s$^{-1}$ for SN\,2020lrt, $\sim 2,800$ km s$^{-1}$ for SN\,2020bpf, and $\sim - 3,100$ km s$^{-1}$ for SN\,2021jun. We are unable to measure a trough of the \ion{Si}{2} line for SN\,2019uqp due to the noisy spectra. Even though the measured velocities are slower than the expected $\sim 10,000$ km s$^{-1}$ from Ia-TDEs, slower velocities within the range we measure are still allowed when accounting for viewing angle effects \citep{MacLeod16}.

\begin{figure}
    \begin{center}
    \centering
    {\includegraphics[width=\columnwidth]{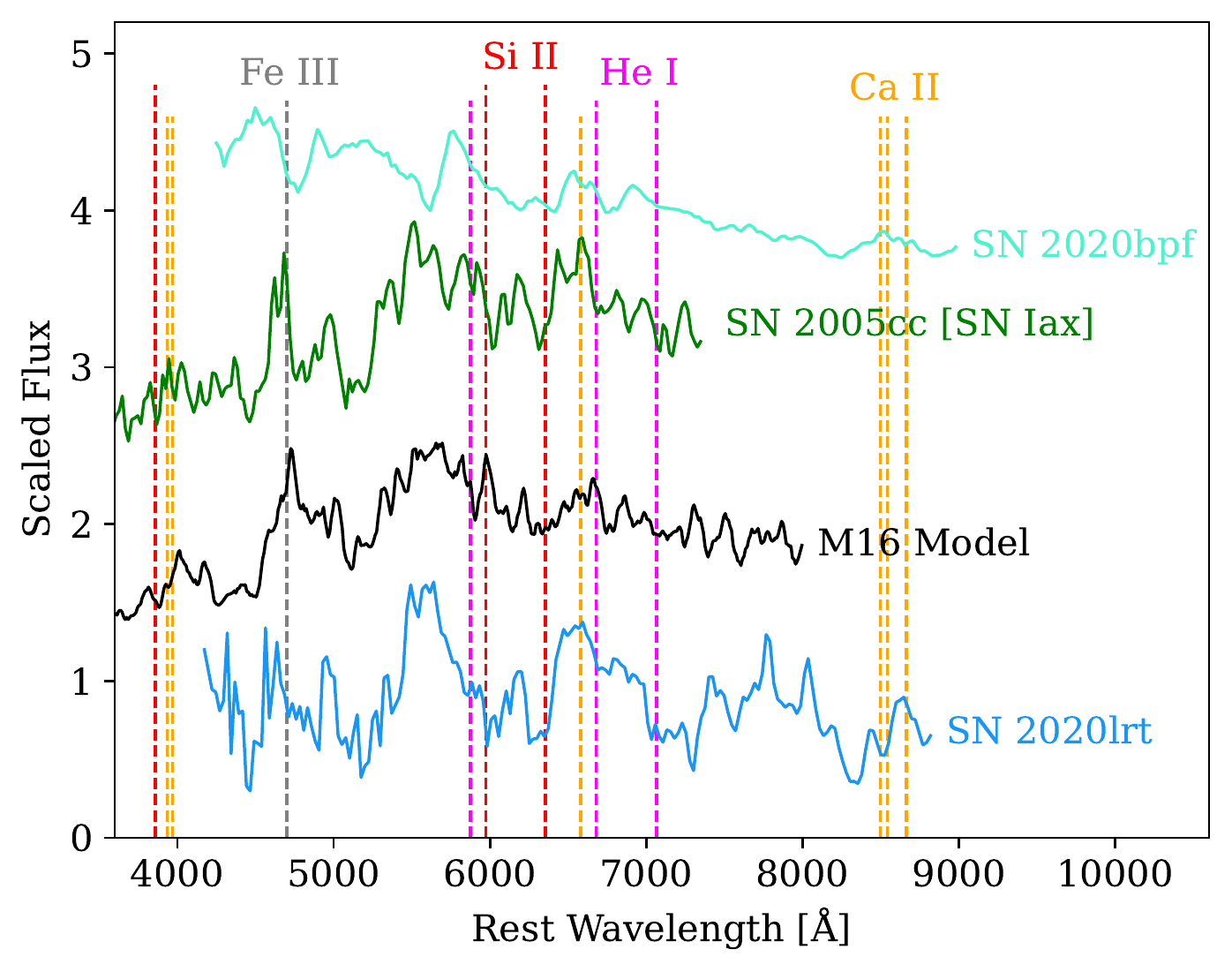}}
    \caption{Spectrum of the most likely Ia-TDE candidate, SN\,2020lrt from \cite{2020lrt}, and a spectrum of SN\,2020bpf classified as a SN Ib by \cite{2020bpf}. For comparison we show the best-fit spectral model from \MM\ to the spectrum of SN\,2020lrt and a comparison SN Iax spectrum from \cite{Blondin12}. We mark common elements as vertical lines. \label{fig:marked}}
    \end{center}
\end{figure}

In Figure~\ref{fig:marked} we show how the Type Iax SN\,2005cc does not show an obvious \ion{Si}{2} velocity offset at a phase of 27 days after explosion \citep{Blondin12}. It is reassuring though to see that for our most likely candidate, SN\,2021lrt, a velocity offset of \ion{Si}{2} is detected at a phase of 31 days. The spectrum of SN\,2021lrt is also an excellent match to the spectral models from \MM. The spectra of both transients and model shown in Figure~\ref{fig:marked} show strong iron absorption around $\sim 5,000$ \AA, \ion{Si}{2} absorption, as well as possible hints of \ion{He}{1}. The SN\,2020lrt spectrum shows emission around $\sim 8,500$ \AA, which could be due to \ion{Ca}{2}. While the \MM\ models and the SN\,2005cc spectrum do not cover this wavelength region, these two have a weaker feature around $\sim 3,900$ \AA, which could similarly be due to \ion{Ca}{2} emission. In conclusion, this makes discerning between Ia-TDE models and Iax spectra particularly challenging. When searching for Ia-TDEs, it is likely they will appear classified as SNe Iax.

\section{Hosts}\label{sec:host}

\begin{figure*}
    \begin{center}
    \centering
    {\includegraphics[width=0.9\textwidth]{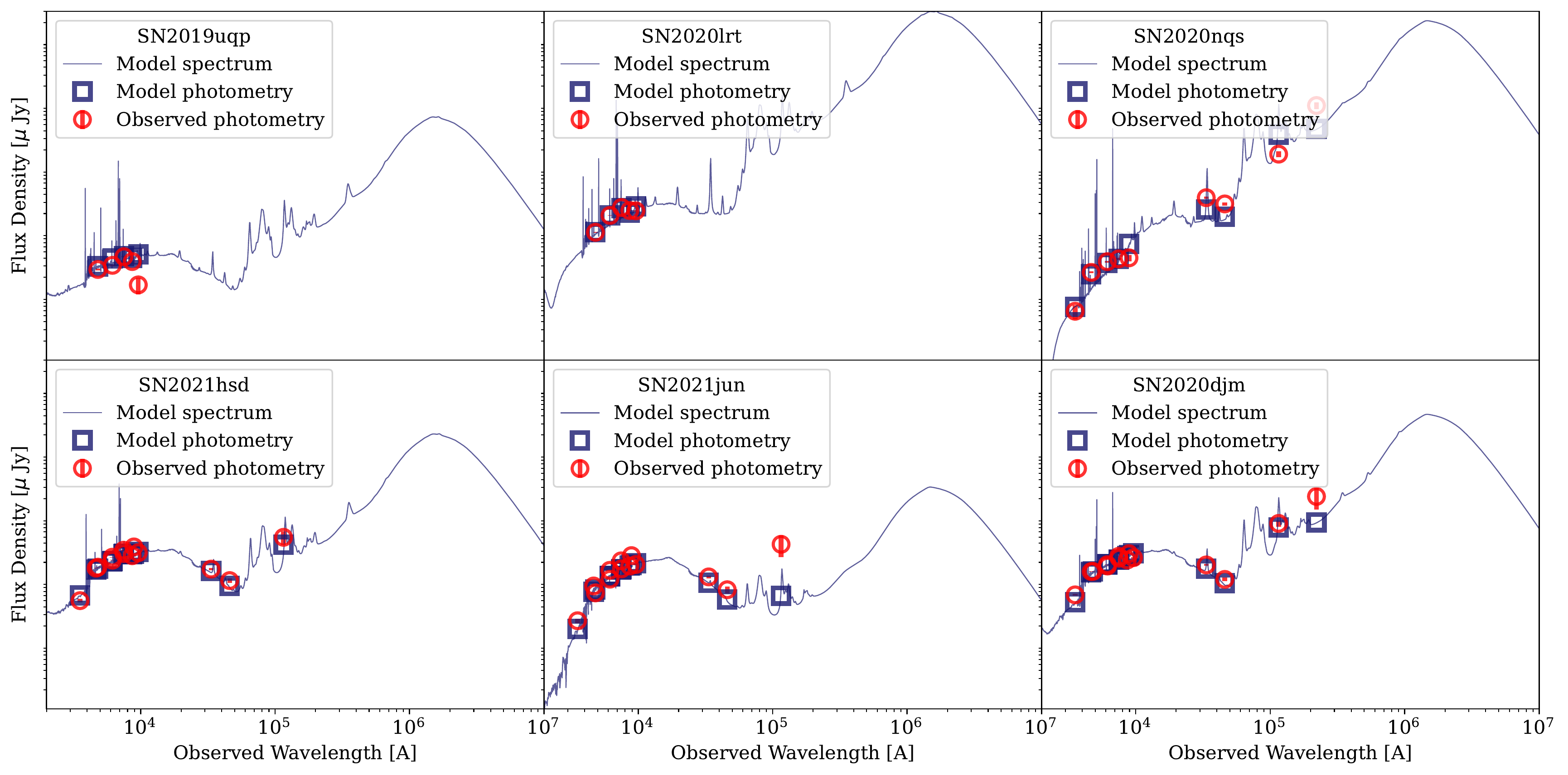}}
    \caption{Photometry of the host galaxies of the nuclear Ia-TDE candidates. The blue line represents the best fit {\tt Prospector} model, and the blue squares are the corresponding model photometry points, with the observed photometry in red. The best-fit stellar mass for each galaxy is shown in Table~\ref{tab:sne}.
    \label{fig:host}}
    \end{center}
\end{figure*}

To characterize the host galaxies of the nuclear Ia-TDE candidates, we model their SEDs using {\tt Prospector}, a software package designed to fit galaxy SEDs by taking into account dust attenuation and re-radiation, nebular emission, and stellar metallicity \citep{leja17}. We fit their spectra using a delay-$\tau$ star formation history (SFH) with an exponential form of the form ${\rm SFR} \propto t^\alpha e ^{-t/\tau}$ \citep{Leja19}. We include photometry from SDSS \citep{Stoughton02}, the PS1/$3\pi$ catalog \citep{Chambers18}, and 2MASS \citep{Skrutskie06} surveys. The resulting SEDs and {\tt Prospector} models of the host galaxies of the nuclear Ia-TDE candidates are shown in Figure~\ref{fig:host}, with their corresponding best fit stellar galaxy mass shown in Table~\ref{tab:sne}. We find that all candidates have a galaxy stellar mass well below the $\log(M_*) < 9.7$ threshold defined by \cite{Besla18} to be considered a dwarf galaxy.

\begin{figure}
    \begin{center}
    \centering
    {\includegraphics[width=\columnwidth]{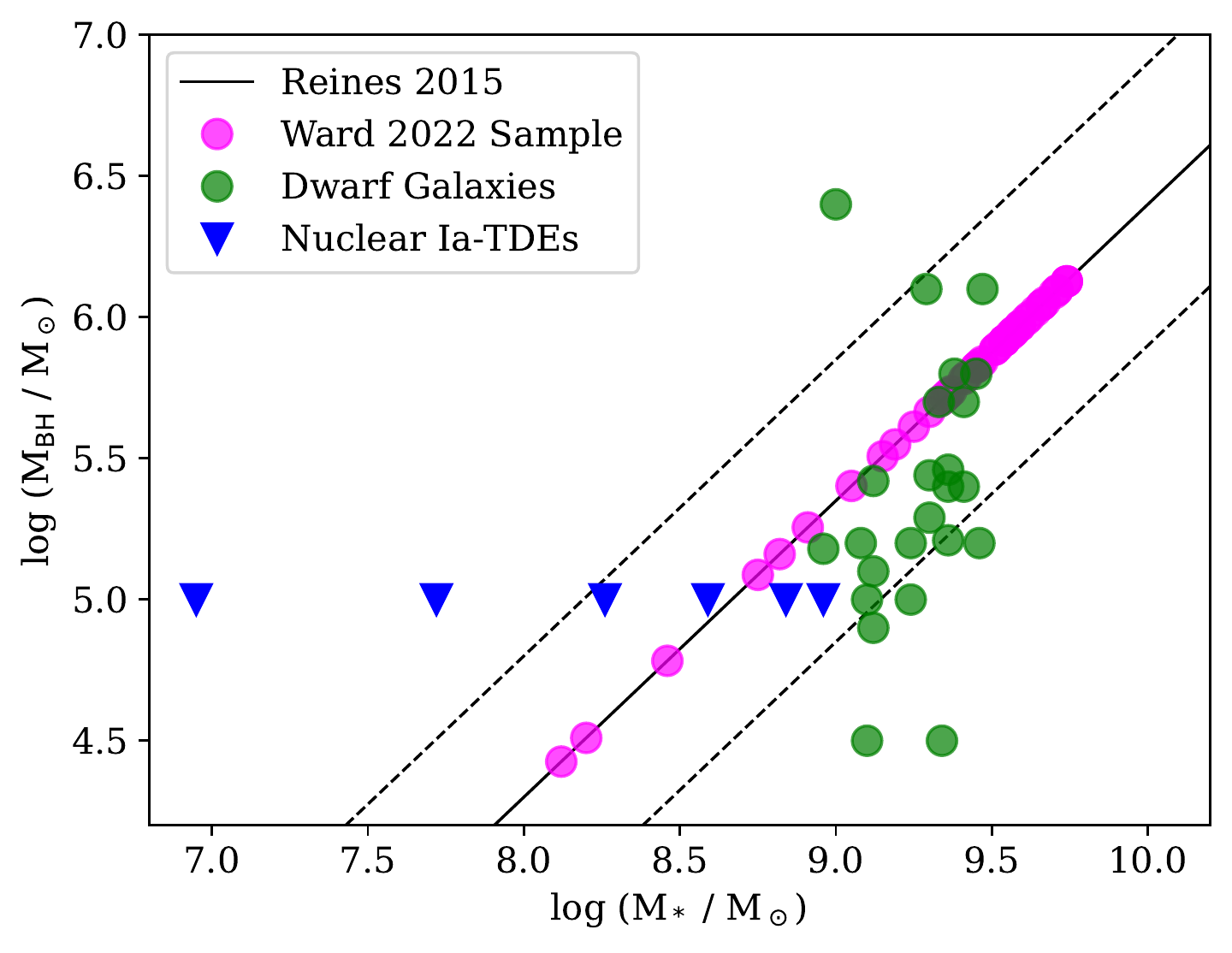}}
    \caption{Black hole mass M$_{\rm BH}$ as a function of galaxy stellar mass M$_*$ for a sample of low mass IMBHs from \cite{Filippenko03, Barth04, Bosch10, Reines13, Baldassare15, Baldassare16, Nguyen18, Nguyen19, Baldassare20}. We also include the M$_*$-M$_{\rm BH}$ relation for dwarf galaxies from \cite{Reines15}, and a sample of dwarf galaxies from \cite{Ward22}, scaled to the \cite{Reines15} relation. In blue we show the best-fit M$_*$ for the Ia-TDE candidates and place them with upper limits on the mass of the BH of $10^5$M$_\odot$. \label{fig:bh_mass}}
    \end{center}
\end{figure}

In Figure~\ref{fig:bh_mass} we show a sample of some of the lowest mass IMBHs known as a function of their host galaxy mass, including the $M_*$-$M_{\rm BH}$ relation for broad-line AGN and dwarf galaxies from \cite{Reines15}. We include the stellar masses of the host galaxies of the nuclear Ia-TDE candidates with the theoretical upper limit for the total mass of their BHs, assuming they are indeed Ia-TDEs. If confirmed to be Ia-TDEs, measuring the masses of their BHs would lead to very strong constraints on the mass function of BHs below $10^5$ M$_\odot$.

The connection between black holes and their galaxies, including the M-$\sigma$ relation, is not well constrained below $10^5$ M$_\odot$. Our current understanding of the mass function of BHs below $10^6$ M$_\odot$ is accurate within only about an order of magnitude \citep{Gallo19}. Current models of supermassive BH formation propose that these can form from initial BH seeds, that grow into IMBHs before reaching their supermassive stage (e.g. \citealt{McKernan12, Belczynski20, Fragione22, Rose22}). Some studies predict that IMBHs do not grow significantly in dwarf galaxies, but instead retain their seed mass for galaxies of a mass up to $\sim 10^{9.5}$ M$_\odot$ \citep{Beckmann22}. Confirming the detection of a Ia-TDE will consequently confirm the discovery of an IMBH $\lesssim 10^5$ M$_\odot$. BHs of masses ranging from $10^5$ to $10^7$ M$_\odot$ have been discovered in galaxies with stellar masses between $10^9$ to $10^{10}$ M$_\odot$ (e.g. \citealt{Pechetti22, Brok15, Nguyen18, Nguyen19, Davis20, Chilingarian18, Mezcua18}), but detections of lower mass BHs remain relatively sparse. AT\,2020neh was one such case, in which the detection of a fast-evolving TDE in a dwarf galaxy implied the tidal disruption of a star by an IMBH of $\sim 10^{4.7-5.9}$ M$_\odot$ \citep{Angus22}, although a direct correlation between the timescale of the optical light curve of a TDE and the central black hole mass is not always evident \citep{Charalampopoulos22}. Finding these TDEs can also help constrain the present-day BH mass function \citep{DOrazio19}, and help bridge the gap between supermassive black holes and stellar-mass black holes (e.g. \citealt{Gultekin14, Arcodia20}).

\section{Discussion}\label{sec:discussion}

\subsection{Future Searches}

\begin{figure}
    \begin{center}
    \centering
    {\includegraphics[width=\columnwidth]{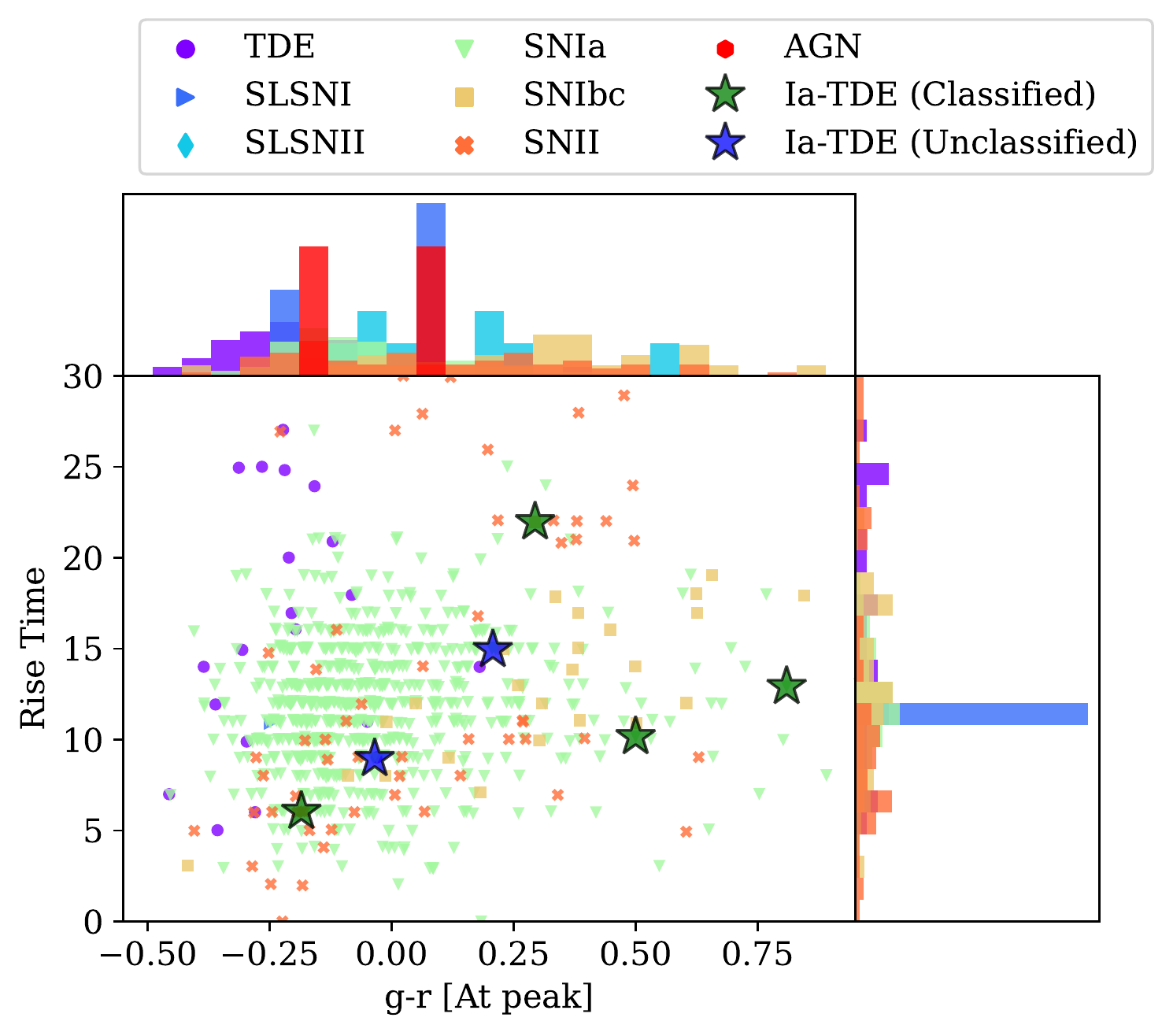}}
    \caption{The $g-r$ color measured during peak compared to the time it takes a transient to rise from discovery to peak for the 1,838 nuclear transients. We also show where the best Ia-TDE candidates lie in this parameter space. \label{fig:histogram}}
    \end{center}
\end{figure}

In this work, we have set the groundwork to systematically search for Ia-TDEs based on their color, rise-time, and relation to their host galaxy. In order to confirm the detection of future Ia-TDEs, we can target follow-up observations of likely candidates before they fade. To do this use plan to use the infrastructure developed for FLEET \citep{Gomez20}. FLEET is a machine learning algorithm originally designed to find superluminous supernovae (SLSNe) or TDEs, which automatically queries the SDSS, PS1/$3\pi$, Gaia, 2MASS, and WISE catalogs, as well as the TNS, and photometry from ZTF and other surveys. We can piggyback off of this infrastructure to select the most likely Ia-TDE candidates based on their observed parameters. In Figure~\ref{fig:histogram} we show an example of two parameters (color and $R_n$) that FLEET uses to select likely SLSNe and TDEs, and how we can use these to select where the Ia-TDE candidates lie in this parameter space.

We test the prospect of following up Ia-TDEs candidates on a list of 6,000 spectroscopically classified transients derived from the original sample described in \S\ref{sec:candidates}, making sure these have a detected host and either a photometric or spectroscopic redshift estimate. For this experiment, we select only transients with $R_n < 0.55$, a host-transient separation $< 0.6\arcsec$, a color during peak $-0.2 < g-r < 0.9$, a color at 40 days past peak of $0.9 < g-r < 3.1$, a host absolute magnitude of $M_r > -19$, and a transient peak absolute magnitude of $M_r > -18$. We find that imposing these cuts returns a total of 36 transients (0.6\% of the initial sample), four of which are among our classified nuclear candidates (2019uqp, 2020lrt, 2020bpf, and 2021jun). This represents an $\sim 11$\% success rate in identifying Ia-TDE candidates from a transient alert stream. We can begin using these selection criteria on ZTF data, but these techniques will be even more relevant once large-scale surveys such as \textit{Rubin} and \textit{Roman} begin in 2024 and 2027, respectively.

The main feature that will allow us to distinguish Ia-TDEs from other optically-similar transients (such as SNe Iax) is the presence of radio emission from a jet afterglow or X-ray emission from the accretion disk formed after the WD-BH encounter. We briefly discuss the feasibility of detecting Ia-TDEs based on the example model described in \MM\ in which a $0.6$ M$_\odot$ WD gets disrupted by a $10^3$ M$_\odot$ BH. In terms of radio emission, VLA observations in the 1 GHz band are able to reach flux limits of $F \sim 100 \mu$Jy. At a typical redshift for the Ia-TDE candidates of $z = 0.04$ (See Table~\ref{tab:sne}), this corresponds to a luminosity of $L \sim 10^{38}$ erg s$^{-1}$. Therefore, based on the model predictions of \MM\, we would be able to detect radio jet afterglows with opening angles $\lesssim 30^\circ$, which are expected to peak at around a month after disruption. For X-ray emission, a typical 3ks \textit{Swift} observation can reach flux limits of $F \sim 10^{-(12-13)}$ erg s$^{-1}$ cm$^{-2}$, which at the same redshift of $z = 0.04$ corresponds to a luminosity of $L \sim 10^{41-42}$ erg s$^{-1}$. Given that the X-ray luminosity from the accretion disk is expected to be Eddington limited at $L \sim 2.5\times10^{42} (M_{\rm BH} / 10^4 M_\odot)$ erg \citep{MacLeod16}, these observations might be marginally able to detect X-ray emission in some Ia-TDEs, but detection will be more likely for the most massive BHs $\gtrsim 10^4$ M$_\odot$.

\subsection{Nuclear Transients}

For this study, we collected a sample of 1,838 classified nuclear transients, before removing the ones with light curves inconsistent with Ia-TDEs. Of these, 73.9\% are SNe Ia, 14.5\% are SNe II, 3.6\% are SNe Ib/c, 2.5\% are TDEs, 2.0\% are AGN, and 3.5\% make up the remaining classes (SLSNe, SNe Ibn, SNe with no subclass, QSO, etc.). The rate of TDEs is higher than that of AGN given that AGN are rarely reported for classification to the TNS. The rate of SNe Ia is comparable to that of the ZTF Bright Transient Survey (BTS) of 71.9\% \citep{Fremling20}. The rate of SNe II and SNe Ib/c are slightly lower than the 20\% and 5\%, respectively reported by the BTS. This suggests a marginal over-representation of SNe Ia in the nuclei of galaxies compared to other classes of nuclear SNe. 

We collected a sample of 813 classified and 630 unclassified nuclear transients. Out of those, we selected four Ia-TDE candidates from the classified sample and three from the unclassified one. This translates to a rate of $< 0.5$\% Ia-TDEs in both the classified and unclassified samples. This is comparable to the total rate of TDEs of $\approx 0.5\%$ in a magnitude-limited survey  \citep{Fremling20}.

\section{Conclusions}\label{sec:conclusions}

We systematically searched the Zwicky Transient Facility alert stream to look for thermonuclear transients from the tidal disruptions of white dwarfs (Ia-TDEs) by comparing these observations to theoretical light curve and spectroscopic models from \cite{MacLeod16}. We restrict our search for nuclear Ia-TDEs to only include transients within $1.5\arcsec$ of their host galaxy, at a normalized radius of $R_n < 1$, with well-characterized non-periodic light curves with at least 5 detections in each $g$- and $r-$ band, with a peak absolute magnitude $M_g \geq -18$, and in galaxies with an absolute magnitude $M_g \geq -19$ to target dwarf galaxies. This resulted in a sample of 46 spectroscopically classified and 90 unclassified transients. From this sample, we found four classified and 2 unclassified transients that match the theoretical predictions from \MM. We find that SN\,2020lrt provides the closest match in terms of both its light curve and spectra to these theoretical model predictions.

Additionally, we perform this same search but without imposing a restriction on the transient having to be nuclear. This selection resulted in a sample of 302 classified and 453 unclassified off-nuclear transients. From this sample, we were able to find only 2 classified and 2 unclassified transients that matched the \MM\ models. This supports the hypothesis that the nuclear candidates are likely Ia-TDEs, as opposed to a transient that is not correlated with being in the nuclei of galaxies.

In order for a BH to disrupt a WD outside the event horizon, the BH needs to be $\lesssim 10^5$M$_\odot$. To date, there have only been a handful of BHs discovered in this mass range, and their mass function below $10^6$M$_\odot$ is poorly understood. Proving that any of these transients is a Ia-TDEs would provide direct evidence for the existence of a BH $\lesssim 10^5$M$_\odot$. Optimizing our transient selection criteria to search for Ia-TDE candidates in existing and future surveys might be the best way we will have to discover more IMBHs, and enable follow-up of their subsequent accretion and/or jet signatures across the electromagnetic spectrum.

\acknowledgements

We thank Morgan MacLeod for re-computing Ia-TDE models in the bands required for this study, and Matthew Siebert for useful discussions regarding the nature of SNe Iax. S.G. is supported by an STScI Postdoctoral Fellowship. This project was supported in part by the Transients Science @ Space Telescope group. Operation of the Pan-STARRS1 telescope is supported by the National Aeronautics and Space Administration under grant No. NNX12AR65G and grant No. NNX14AM74G issued through the NEO Observation Program. The ZTF forced-photometry service was funded under the Heising-Simons Foundation grant \#12540303 (PI: Graham). This work has made use of data from the European Space Agency (ESA) mission {\it Gaia} (\url{https://www.cosmos.esa.int/gaia}), processed by the {\it Gaia} Data Processing and Analysis Consortium (DPAC, \url{https://www.cosmos.esa.int/web/gaia/dpac/consortium}). Funding for the DPAC has been provided by national institutions, in particular the institutions participating in the {\it Gaia} Multilateral Agreement. This research has made use of NASA’s Astrophysics Data System. This research has made use of the SIMBAD database, operated at CDS, Strasbourg, France. This research has made use of the NASA/IPAC Extragalactic Database, which is funded by the National Aeronautics and Space Administration and operated by the California Institute of Technology. This work makes use of the Weizmann Interactive Supernova Data Repository (WISeREP, \url{https://www.wiserep.org}). 

\facilities{ADS, Gaia, OSC, MAST, PS1, Sloan, TNS, WISE, 2MASS}
\software{Astropy \citep{astropy}, dynesty \citep{dynesty}, extinction \citep{Barbary16}, FLEET \citep{Gomez20_FLEET}, Matplotlib \citep{matplotlib}, mpi4py \citep{mpi4py}, NumPy \citep{numpy}, Prospector \citep{leja17}, PyMC \citep{pymc3}, schwimmbad \citep{schwimmbad17}, SciPy \citep{Scipy}, sedpy \citep{sedpy}, Superphot \citep{Hosseinzadeh20}}

\bibliography{references}

\end{document}